\documentclass[oneside]{article}
\usepackage{graphicx}
\usepackage{subcaption}
\usepackage[final]{nips_2018}
\usepackage{graphicx}
\usepackage{amsmath, amssymb}

\usepackage{amsthm}
 
\theoremstyle{definition}

\newcommand{\indep}{\rotatebox[origin=c]{90}{$\models$}}
\renewcommand{\bibname}{References}
\renewcommand{\bibsection}{\subsubsection*{\bibname}}

\begin{document}

%

%

\title{Estimating Causal Effects With Partial Covariates For Clinical Interpretability}

\author{
  Sonali Parbhoo\thanks{These authors contributed equally.},\hspace{0.015cm} Mario Wieser\footnotemark[1] \hspace{0.1cm}and Volker Roth\\
  University of Basel\\
  \texttt{name.surname@unibas.ch} \\
}
\maketitle
\section{Introduction}
Understanding the causal effects of an intervention is a key question in many applications, from personalised medicine to marketing (e.g. \cite{sun2015causal,wager2017estimation, AlaaS17}). Predicting the causal outcome typically involves dealing with high-dimensional observational data that is frequently subject to the effects of \emph{confounding}.

In general, we distinguish between measured and hidden confounding: When confounders are directly measured, they may be accounted for using techniques that correct for their effects, such as propensity reweighting (IPS) or covariate shift \citep{hernan2006estimating, rosenbaum1984reducing}. In contrast, to account for hidden confounding, proxy variables may be used as noisy representatives of latent confounders \citep{Greenland08, pearl2012measurement, kuroki2014measurement, Louizos}.  Both approaches can however only be applied when covariate data is completely measured. This assumption is not feasible in a large number of settings such as medicine. For example, doctors are interested in identifying treatments that improve patient outcomes, and have to base decisions on hundreds of potentially confounding variables such as age and genetic factors. Here, a doctor may readily have access to many routine measurements such as blood count data for all patients, but may only have genetic information for some patients. Inferring the causal effects of a treatment requires learning a joint distribution over covariates and confounders of patients whose data is completely observable, while simultaneously transferring this knowledge to patients whose data is missing. This is not achievable in practice since we have to integrate over all missing covariates. 

We propose addressing the problem of performing causal inference with partial covariate information from an decision-theoretic point of view. Specifically, we assume that a fixed set of measurements is unavailable for a subset of the data (or patients) at test time. The key idea is to use the Information Bottleneck (IB) criterion \citep{art:tishby:ib} to perform a sufficient reduction of the covariate and recover a distribution of the confounding information. The IB enables us to build a discrete reference class over patients whose covariate data is complete, to which we can map patients with incomplete data and estimate treatment effects on the basis of such a mapping. Finally, we demonstrate that our method outperforms existing approaches across established causal inference benchmarks and a real world application for treating sepsis. 

\section{Method}
\label{sec:method}
\begin{figure}[!h]
\begin{center}
\includegraphics[width=0.50\textwidth]{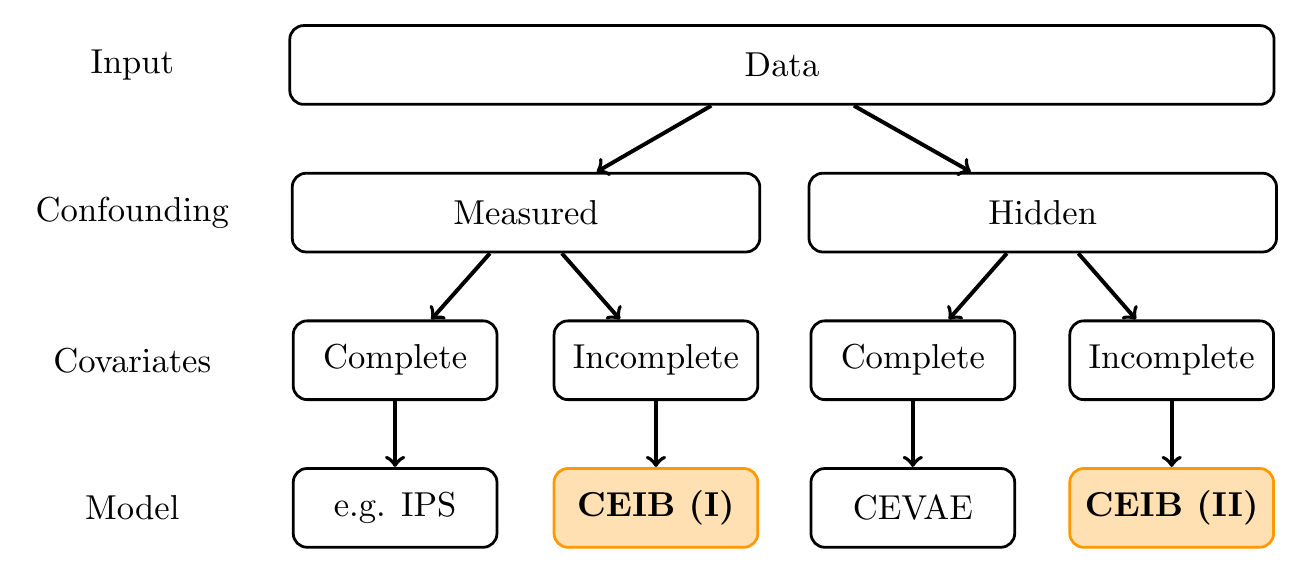}
\caption{Overview of causal inference with confounding effects and missing covariates. In this paper, we address Cases I and II, thus accounting for incomplete covariate information when confounding is measured and hidden respectively.} 
\label{fig:bigpicture}
\end{center}
\end{figure}

\begin{figure*}[!h]
    \centering
    \begin{subfigure}{0.35\textwidth}
        \centering
        \includegraphics{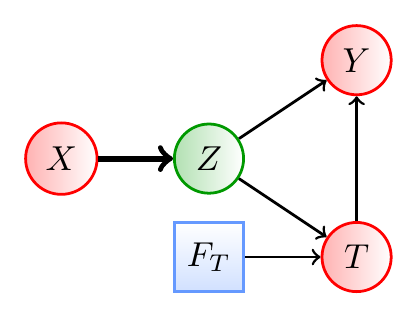}
        \caption{Case I: Measured confounding}
    \end{subfigure}%
    \hspace{2.5cm}
    \begin{subfigure}{0.35\textwidth}
        \centering
        \includegraphics{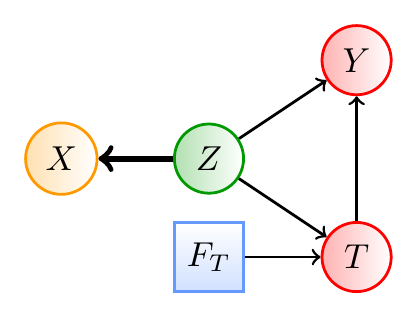}
        \caption{Case II: Hidden confounding}
        \label{figure:case2}
    \end{subfigure}
    \hspace{2.5cm}
    \caption{\label{fig:cases}Influence diagrams of the two cases considered in this paper. Red and green circles correspond to observed and latent random variables respectively, while blue rectangles represent interventions. In Case I, we identify a low-dimensional representation $Z$ of \emph{measured} covariates $X$ to estimate the effects of an intervention on outcome $Y$. In Case II, the arrow between $X$ and $Z$ is \emph{reversed} and confounders are indirectly measured via proxy variables, indicated by an orange circle here. We identify a low-dimensional representation $Z$ and use this to explicitly estimate $X$ as well as $Y$. In both cases, representation $Z$ is used to make inferences for a subset of patients where only partial covariate information is available.}
\end{figure*}

We refer to our model as cause-effect IB (CEIB). In Figure \ref{fig:bigpicture}, we illustrate an overview of the possible configurations for performing causal inference and present our model in the context of existing work. The corresponding causal graphs for Cases I and II are shown in Figure \ref{fig:cases}. The major difference between I and II is the reversal of the arrow between $Z$ and $X$, and the fact that in Case II confounders are not measured, but indirectly observed via noisy proxies.

In our paper, we consider the decision-theoretic approach of \cite{dawid07} to estimate the causal effect where we have both hidden and measured confounding with incomplete covariates. This involves computing the ACE of $T$ on $Y$ .  \citet{dawid07} show that the ACE and observational ACE are equivalent under the conditional independence assumption $Y \indep F_T \mid T$. This assumption expresses that the distribution of $Y \mid T$ is the same in the interventional and observational regimes. It can also be extended to account for the notion of confounding. Here, the treatment assignment $F_T$ may be ignored when estimating $Y$, provided a sufficient covariate $Z$ and $T$. Formally, $Z$ is a sufficient covariate for the effect of $T$ on outcome $Y$ if $Z \indep  F_T$ and $Y \indep F_T \mid (Z,T)$. It can also be shown via Pearl's backdoor criterion \citep{pearl2009causality} that the ACE may be defined in terms of the Specific Causal Effect (SCE),
\begin{align} \label{eq5}
\begin{split}
   ACE := \mathbb{E}[SCE(Z, F_T=\varnothing)]
\end{split}
\end{align}
where
\begin{align} \label{eq6}
\begin{split}
   SCE(X) :&= \mathbb{E}[Y \mid Z, T=1, F_T=\varnothing)] \\&- \mathbb{E}[Y \mid Z, T=0, F_T=\varnothing)].
\end{split}
\end{align}
Importantly, estimating the ACE only requires computing a distribution $Y|Z, T$ in Figure \ref{fig:cases}. In what follows, we use the IB to learn a sufficient covariate that allows us to approximate this distribution.

\paragraph{Case I: Measured Confounding}
This case occurs when we have observational data where all the relevant confounding variables are measured, but where a fixed set of covariates is only available for some subset of the data at test time. Let $X_1$ and $X_2$ be our covariate sets (both available at training). We adapt the IB for learning the outcome of a therapy when partial covariate information is available for $X_2$ at test time. To do so, we consider the following parametric form, 
\begin{align}
\max_{\phi, \theta, \psi, \eta} -I_{\phi} (V_1; X_1) - I_{\eta}(V_2; X_2) + \lambda I_{\phi, \theta, \psi, \eta}(Z; (Y, T)), 
\label{criterion}
\end{align}
where $V_1$ and $V_2$ are low-dimensional discrete representations of the covariate data, $Z = (V_1, V_2)$ is a concatenation of $V_1$ and $V_2$ and $I$ represents the mutual information parameterised by networks $\phi$, $\psi$, $\theta$ and $\eta$ respectively. We assume a parametric form of the conditionals $q_\phi(v_1|x)$, $q_\eta(v_2|x)$, $p_\theta(y|t, z)$, $p_\psi(t|z)$, as well as Markov chain $Z - X - T - Y$. The three terms in Equation \ref{criterion} have the following forms:
\begin{gather}
\begin{align}
I_\phi(V_1; X_1) &= D_{KL}(q_\phi(v_1|x_1)p(x_1)||p(v_1)p(x_1)) \nonumber\\ &= \mathbb{E}_{p(x_1)}D_{KL}(q_\phi(v_1|x_1)||p(v_1)) \\
I_\eta(V_2; X_2) &= D_{KL}(q_\eta(v_2|x_2)p(x_2)||p(v_2)p(x_2)) \nonumber\\ &= \mathbb{E}_{p(x_2)}D_{KL}(q_\eta(v_2|x_2)||p(v_2)) \\
I_{\phi, \theta, \psi, \eta}(Z; (Y, T)) 
&= \mathbb{E}_{p(x, y, t)} \mathbb{E}_{p_{\phi,\eta}(z|x)} \log p_\theta(y|t, z)  \nonumber\\&+ \log p_\psi(t|z) + h(y), 
\end{align}
\end{gather}
as a result of the Markov assumption in the IB model. Here $h(y) = -\mathbb{E}_{p(y)} \log p(y)$ is the entropy of $y$. For the decoder model, we use an architecture similar to the TARnet \citep{JohanssonTAR}, where we replace conditioning on high-dimensional covariates $X$ with conditioning on latent $Z$. We can thus express the conditionals as, 
\begin{align}
p_\psi(t|z) = \textup{Bern}(\sigma(f_1(z)))\hspace{0.5cm} \nonumber\\
p_\theta(y|t, z) = \mathcal{N}(\mu = \hat\mu, \sigma^2 = \hat{s}),
\end{align}
with logistic function $\sigma(\cdot)$, and outcome $Y$ given by a Gaussian distribution parameterised with a TARnet with $\hat\mu = t f_2(z) + (1- t)f_3(z)$. Note that the terms $f_k$ correspond to neural networks.  

\paragraph{Case II: Hidden Confounding}
\label{sec:hiddenconv}
This case is analogous to the work of \citet{Louizos}. We however, treat proxies as measured confounders and propose using Case I to estimate the causal effect here. Using Case I is permissible since both DAGs in Figure \ref{fig:cases} are Markov equivalent, and the causal direction between $X$ and $Z$ can only be determined by additional assumptions on the causal graph. However, assuming the causal structure in Figure \ref{figure:case2} as in \citet{Louizos} requires the definition of a complex prior over $Z$. Hence, it may be more natural to treat all covariates including proxies as measured confounders like we propose in this paper. In doing so, we compress the relevant information to a sufficient covariate as described in Case I. 

Once we can estimate $y$ in both cases using the proposed model, we can compute the ACE. When given a test patient with partial covariates, we can assign them to the closest equivalence class of patients with similar characteristics, and approximate the effect of treatments this basis.

\section{Experiments}
\label{sec:experiments}
We demonstrate the performance of our approach on a high-dimensional real world task for managing and treating sepsis. Additional experiments are in the supplement. For this experiment, we make use of data from the Multiparameter Intelligent Monitoring in Intensive Care (MIMIC-III) database \citep{johnson2016mimic}. We focus on patients satisfying Sepsis-3 criteria (16 804 patients in total). For each patient, we have a 48-dimensional set of physiological parameters including demographics, lab values, vital signs and input/output events, where covariates are partially incomplete.  Our outcomes $y$ correspond to the odds of mortality, while we binarise medical interventions $t$ according to whether or not a vasopressor is administered.  The data set is divided into 60/20/20\% into training/validation/testing sets. We train our model with 6, 4-dimensional Gaussian mixture components and analysed the information curves and cluster compositions respectively. 

\begin{figure}[!h]
    \centering
        \begin{subfigure}{0.45\textwidth}
        		\centering
        		\includegraphics[width=0.8\textwidth]{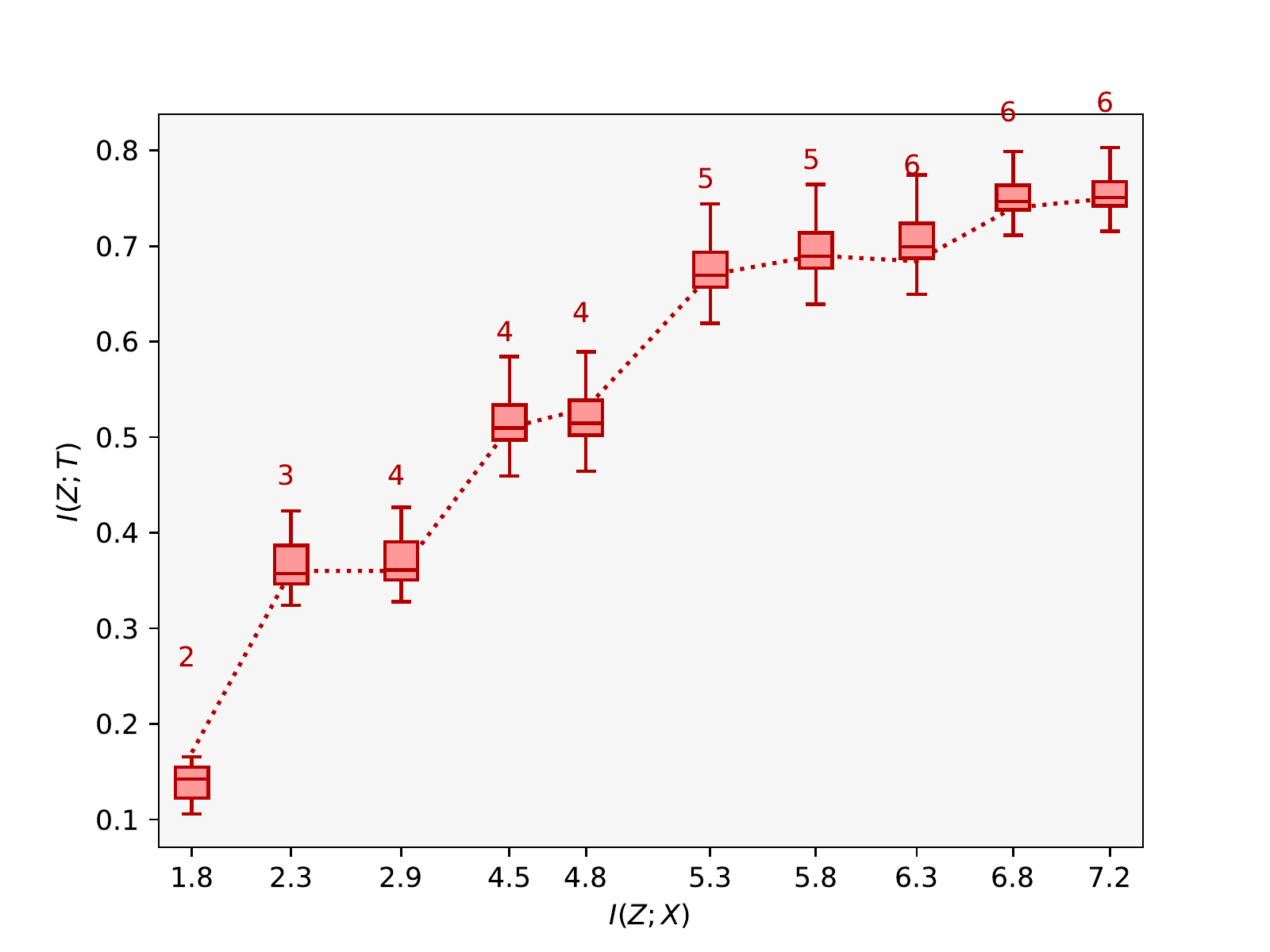}
        		\caption{}
        		\label{information curve:c}
        \end{subfigure}         
        \begin{subfigure}{0.45\textwidth}
                \centering
        		\includegraphics[width=0.8\textwidth]{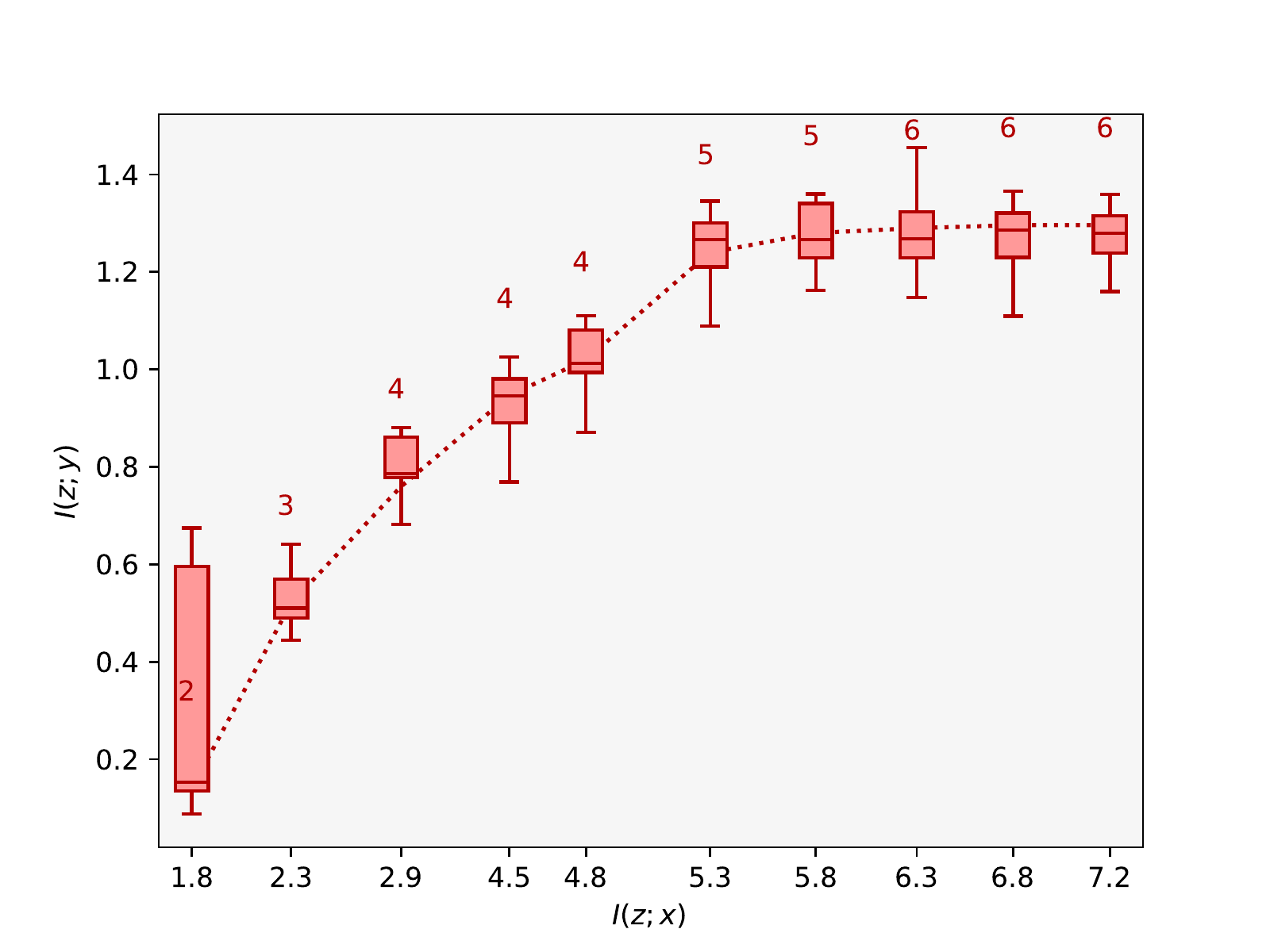}
        		\caption{}
        		\label{information curve:d}
        \end{subfigure}  
    \caption{Subfigures (a) and (b) illustrate the information curve $I(Z; T)$ and $I(Z; Y)$ for the task of managing sepsis. We perform a sufficient reduction of the covariates to 6-dimensions and are able to approximate the ACE on the basis of this.}  
\end{figure}   

\begin{figure*}[!h]
    \centering
        \begin{subfigure}{0.15\textwidth}
        		\centering
        		\includegraphics[width=\textwidth]{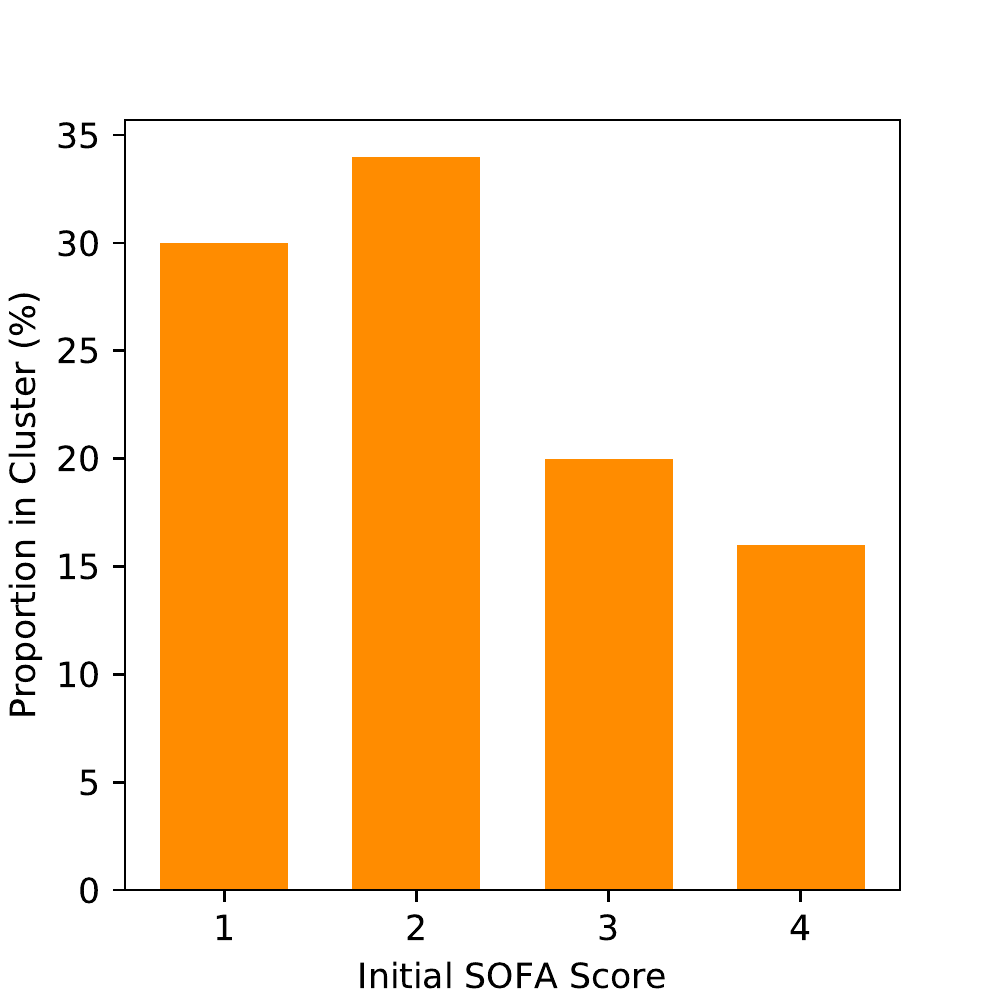}
        		\caption{Cluster 1}
        		\label{cluster1s}
        \end{subfigure}         
        \begin{subfigure}{0.15\textwidth}
        		\centering
        		\includegraphics[width=\textwidth]{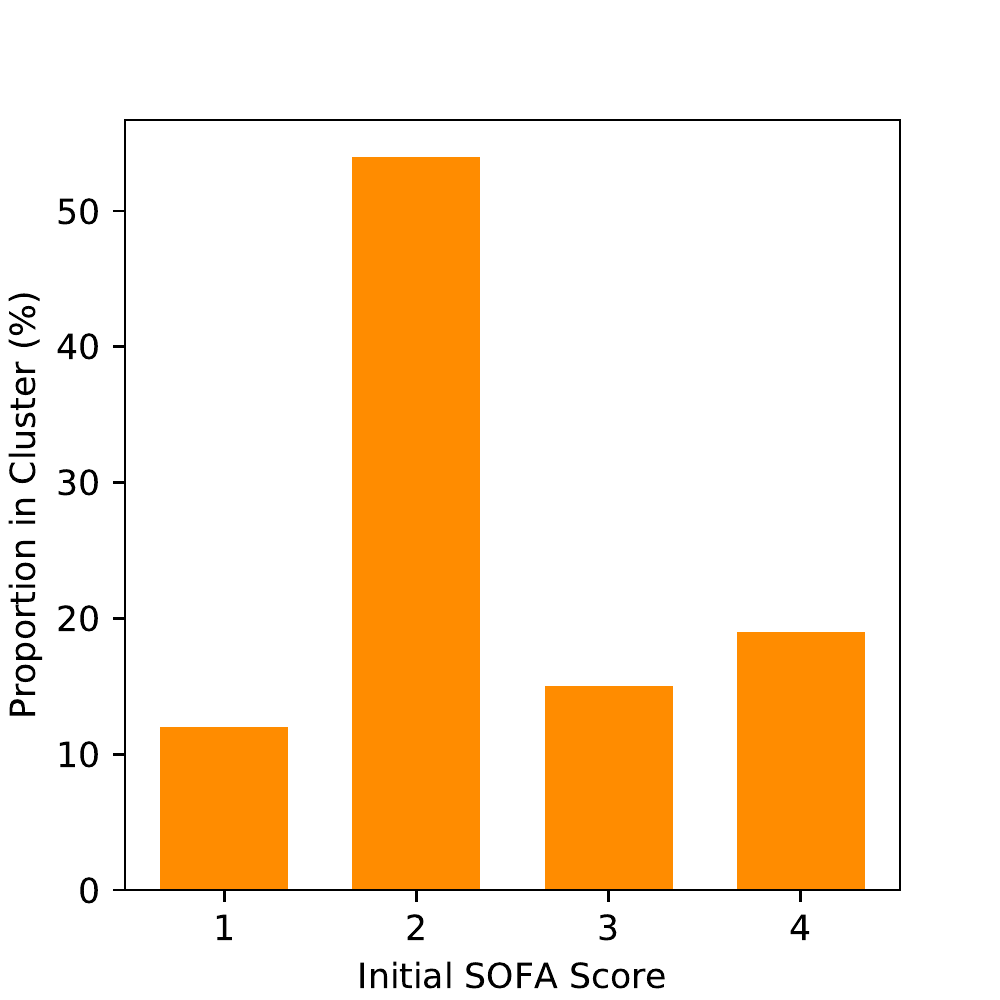}
        		\caption{Cluster 2}
        		\label{cluster2s}
        \end{subfigure}
        \begin{subfigure}{0.15\textwidth}
                \centering
        		\includegraphics[width=\textwidth]{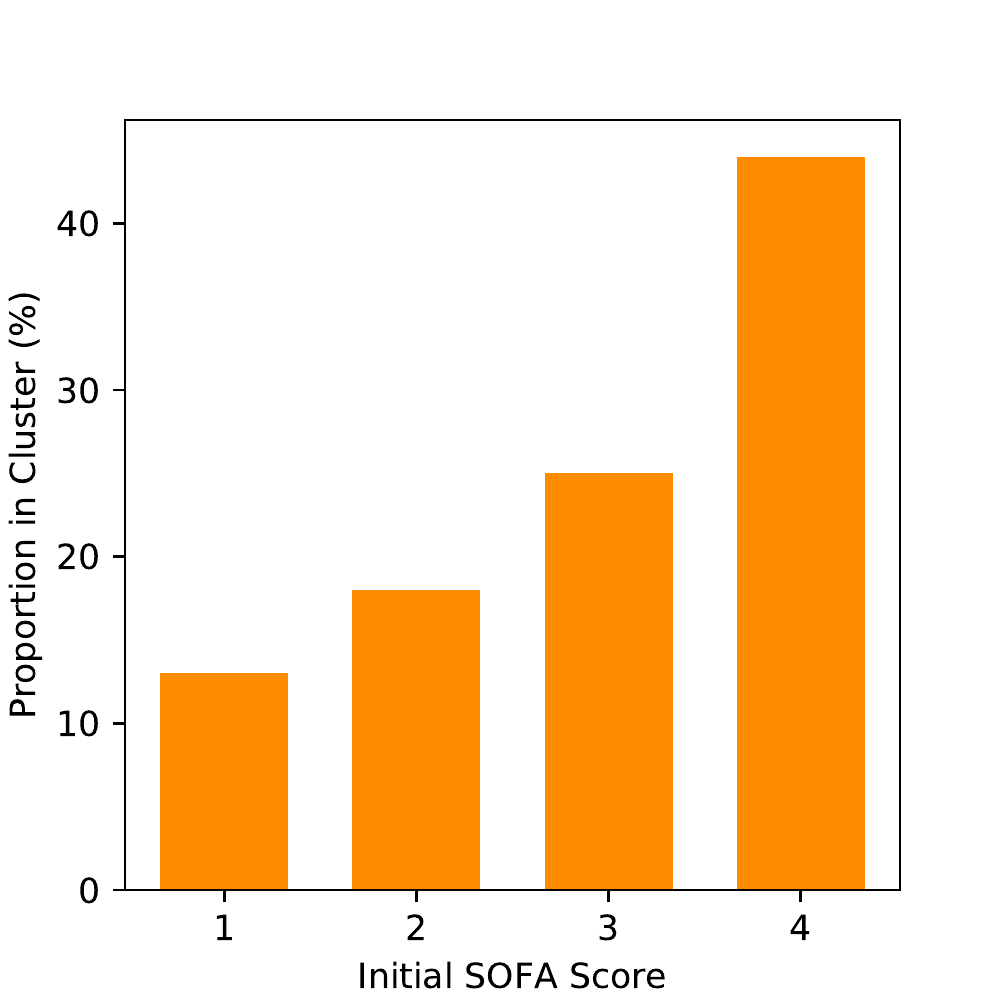}
        		\caption{Cluster 3}
        		\label{cluster3s}
        \end{subfigure}  
             \begin{subfigure}{0.15\textwidth}
                \centering
        		\includegraphics[width=\textwidth]{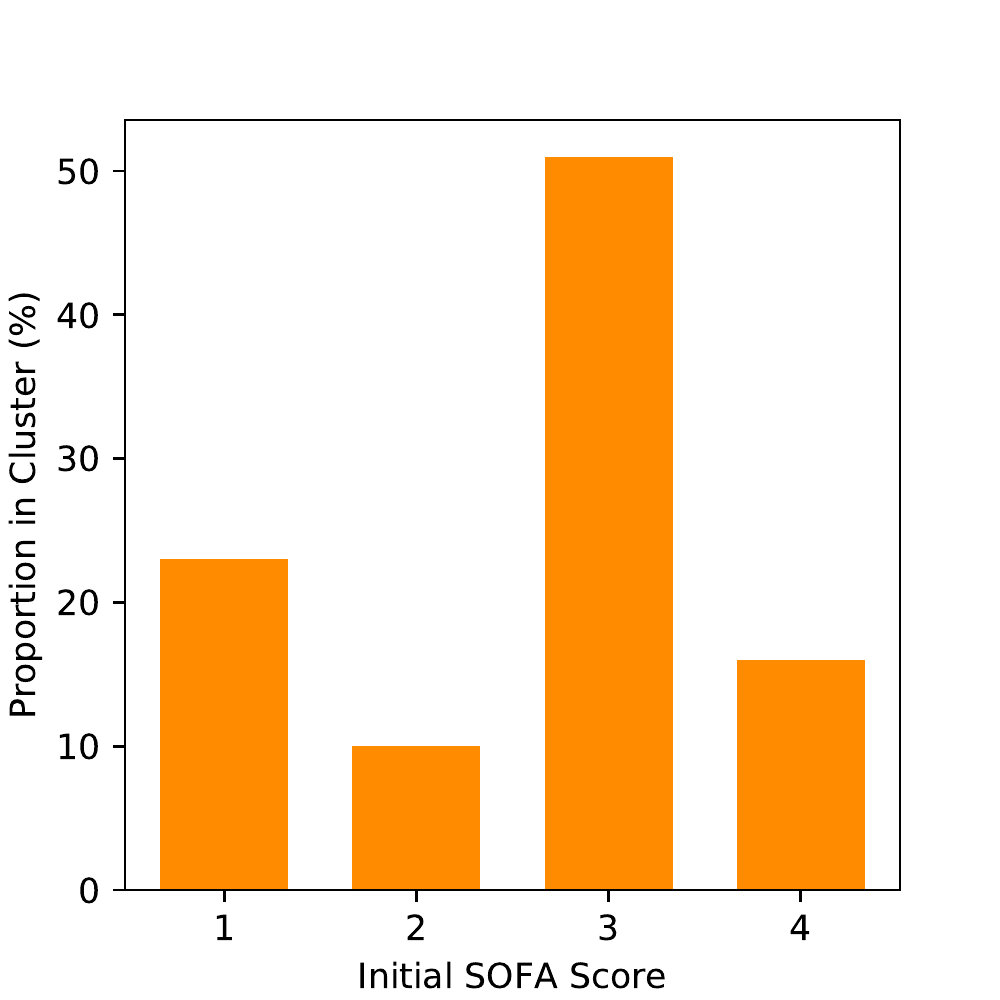}
        		\caption{Cluster 4}
        		\label{cluster4s}
		\end{subfigure}  
		  \begin{subfigure}{0.15\textwidth}
                \centering
        		\includegraphics[width=\textwidth]{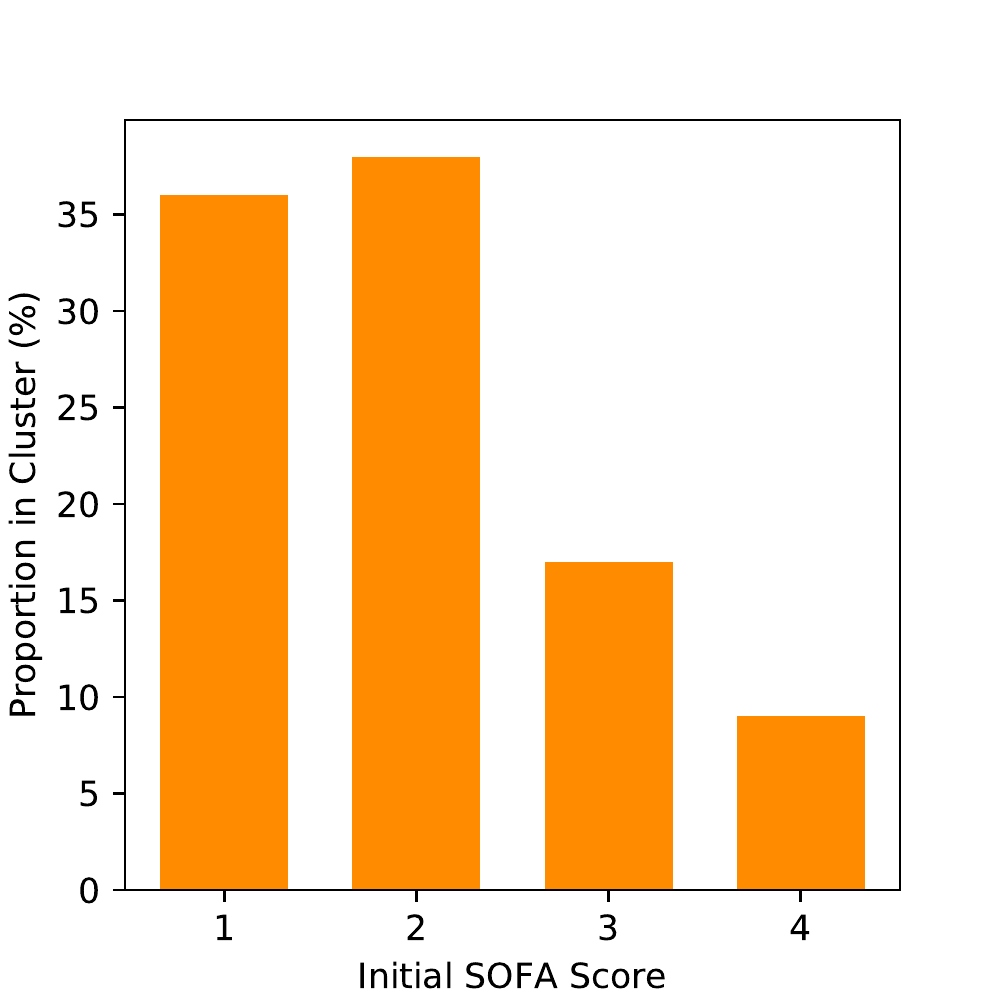}
        		\caption{Cluster 5}
        		\label{cluster5s}
		\end{subfigure}  
		  \begin{subfigure}{0.15\textwidth}
                \centering
        		\includegraphics[width=\textwidth]{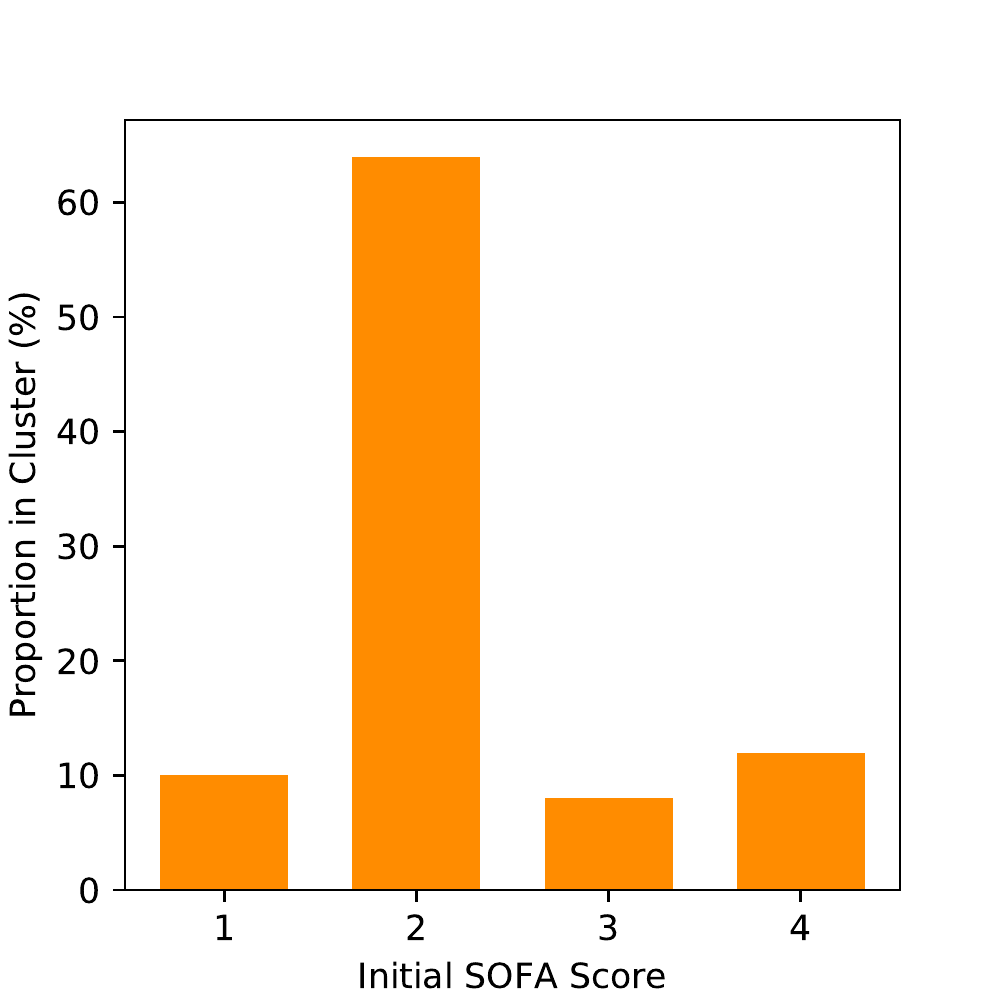}
        		\caption{Cluster 6}
        		\label{cluster6s}
		\end{subfigure}  
		\caption{\label{sepsis clusters} Proportion of initial SOFA scores in each cluster. The variation in initial SOFA scores across clusters suggests that it is a potential confounder of odds of mortality when managing and treating sepsis.} 
\label{fig:sepsis-clusters}
\end{figure*}

The information curves for $I(Z; T)$ and $I(Z; Y)$ are shown in Figures \ref{information curve:c} and \ref{information curve:d} respectively. We observe that we can perform a sufficient reduction of the high-dimensional covariate information to between 4 and 6 dimensions while achieving high predictive accuracy of outcomes $y$. Since there is no ground truth available for the sepsis task, we do not have access to the true confounding variables. However, we can perform an analysis on the basis of the clusters obtained over the latent space. Here, we see that we can characterise the patients in each cluster according to their initial SOFA (Sequential Organ Failure Assessment) scores.  SOFA scores range between 1-4 and are used to track a patient's stay in hospital. In Figure \ref{fig:sepsis-clusters}, we observe clear differences in cluster composition relative to the SOFA scores.  Clusters 2, 5 and 6 tend to have higher proportions of patients with lower SOFA scores, while Clusters 3 and 4 have larger proportions of patients with higher SOFA scores. This result suggests that a patient's initial SOFA score is potentially a confounder when determining how to administer subsequent treatments and  predicting their odds of in-hospital mortality. This is consistent with medical studies such as \citet{medam17, studnek2012impact} where authors indicate that high initial SOFA scores were likely to impact on their overall chances of survival and treatments administered in hospital. Overall, performing such analyses for tasks like Sepsis may help correct for confounding and assist in establishing potential guidelines.

\section{Discussion}
\label{discussion}
\paragraph{CEIB makes state-of-the-art predictions of the ACE that are robust against confounding.}
\paragraph{CEIB learns a low-dimensional, interpretable representation of latent confounding.}
Since CEIB extracts only the information that is relevant for making predictions, it is able to learn a \emph{low-dimensional} representation of the confounding effect and uses this to make predictions. In particular, the introduction of a discrete cluster structure in the latent space allows an easier interpretation of the confounding effect. Similar methods such as \citet{Louizos} typically use a higher dimensional representation to account for these effects without gains in performance.  This is likely a result of misrepresenting the true confounding effect. Modelling the task as an IB alleviates this problem. For sepsis, we identify a latent space of 6 dimensions when predicting odds of mortality, where clusters exhibit a distinct structure with respect to a patient's initial SOFA score. 
\paragraph{CEIB enables estimating the causal effect with incomplete covariates.}
Unlike previous approaches, CEIB can deal with incomplete covariate data during test time by introducing a discrete latent space. Specifically, we learn equivalence classes among patients such that the approximate the effects of treatments can be computed where data is incomplete.

\bibliography{paper}
\bibliographystyle{plainnat}
\newpage

\appendix

\section{Additional Experiments}
\subsection{Infant Health and Development Program}
The Infant Health and Development Program (IHDP) \citep{IHDP, JHill} is a randomised control experiment assessing the impact of educational intervention on outcomes of pre-mature, low birth weight infants born in 1984-1985. Measurements from children and their mother were collected for studying the effects of childcare and home visits from a trained specialist on test scores.  Briefly, the study contains information about the children and their mothers/caregivers. Data on the children include treatment group, sex, birth weight, health indices. Information about the mothers includes maternal age, mother's race as well as educational achievement. \citet{JHill} extract features and treatment assignments from the real-world clinical trial, and introduce selection bias to the data artificially by removing a non-random portion of the treatment group, in particular children with non-white mothers. In total, the data set consists of 747 subjects (139 treated, 608 control), each represented by 25 covariates measuring properties of the child and their mother. The data set is divided into 60/20/20\% into training/validation/testing sets.

For our experiments, we compare the performance of CEIB for predicting the ACE against several existing baselines as in \citet{Louizos}: OLS-1 is a least squares regression; OLS-2 uses two separate least squares regressions to fit the treatment and control groups respectively; TARnet is a feedforward neural network from \citet{pmlr-v70-shalit17a}; KNN is a $k$-nearest neighbours regression; RF is a random forest; BNN is a balancing neural network \citep{JohanssonTAR}; BLR is a balancing linear regression \citep{JohanssonTAR}, and CFRW is a counterfactual regression that using the Wasserstein distance \citep{pmlr-v70-shalit17a}. 

\begin{table}[!htbp]
  \centering
  \begin{tabular}{lclc|c|}
    \hline
    Method  & $\epsilon_{ACE}^{within-s}$  & $\epsilon_{ACE}^{out-of-s}$\\
    \hline
    OLS-1  & $.73\pm.04$  & $.94\pm.06$ \\
    OLS-2 &  $\mathbf{.14\pm.01}$  & $.31\pm.02$ \\
    KNN & $\mathbf{.14\pm.01}$  & $.79\pm.05$  \\
    BLR &  $.72\pm.04$  & $.93\pm.05$\\
    TARnet & $.26\pm.01$ & $.28\pm.01$ \\
    BNN  & $.37\pm.03$ & $.42\pm.03$ \\
    RF  & $.73\pm.05$  & $.96\pm.06$ \\
    CEVAE  & $.34\pm.01$  & $.46\pm.02$ \\
    CFRW  & $.25\pm.01$ & $.27\pm.01$\\
    \hline
    CEIB  & $.15\pm.02$  & $\mathbf{.23\pm.01}$ \\
    \hline
  \end{tabular}
  \caption{ \label{table:ihdp}Within-sample and out-of-sample mean and standard errors for the metrics across models on the IHDP data set. A smaller value indicates better performance. Bold values indicate the method with the best performance. }
\end{table}

We train our model with $k = 4$, $d = 3$-dimensional Gaussian mixture components, although our method can be applied without loss of generality to any number of dimensions. To assess the ability to estimate treatment effects on the basis of partial information, we artificially exclude three covariates at test time. These are covariates that are exhibit a moderate correlation to the hidden confounder ethnicity. The results are shown in Table \ref{table:ihdp}. Overall, our approach exhibits good performance for both in-sample and out-of-sample predictions, while simultaneously accounting for partial covariate information.

\begin{figure}[!h]
    \centering
        \begin{subfigure}{0.45\textwidth}
        		\centering
        		\includegraphics[width=0.8\textwidth]{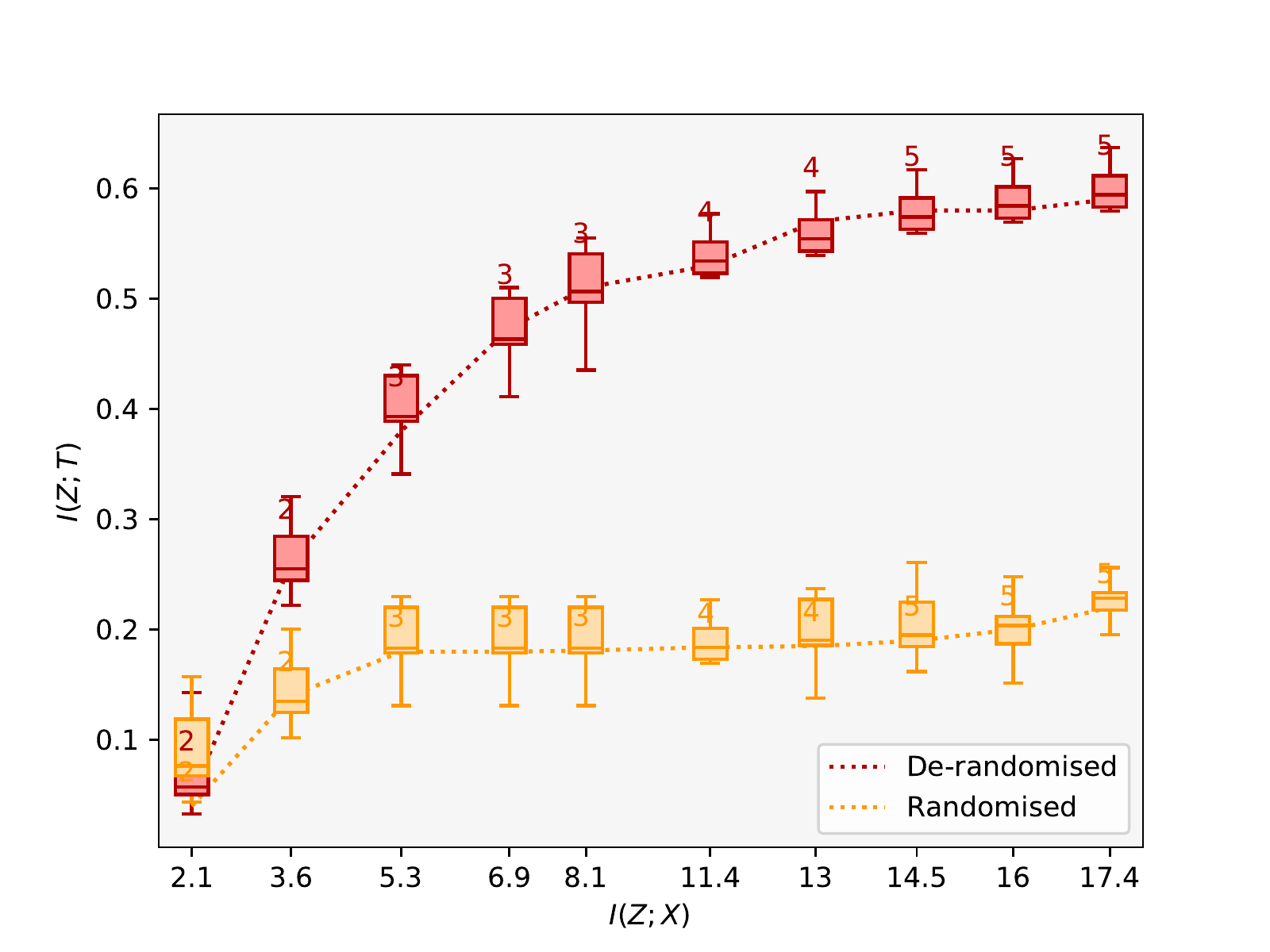}
        		\caption{}
        		\label{information curve:b}
        \end{subfigure}         
        \begin{subfigure}{0.45\textwidth}
                \centering
        		\includegraphics[width=0.8\textwidth]{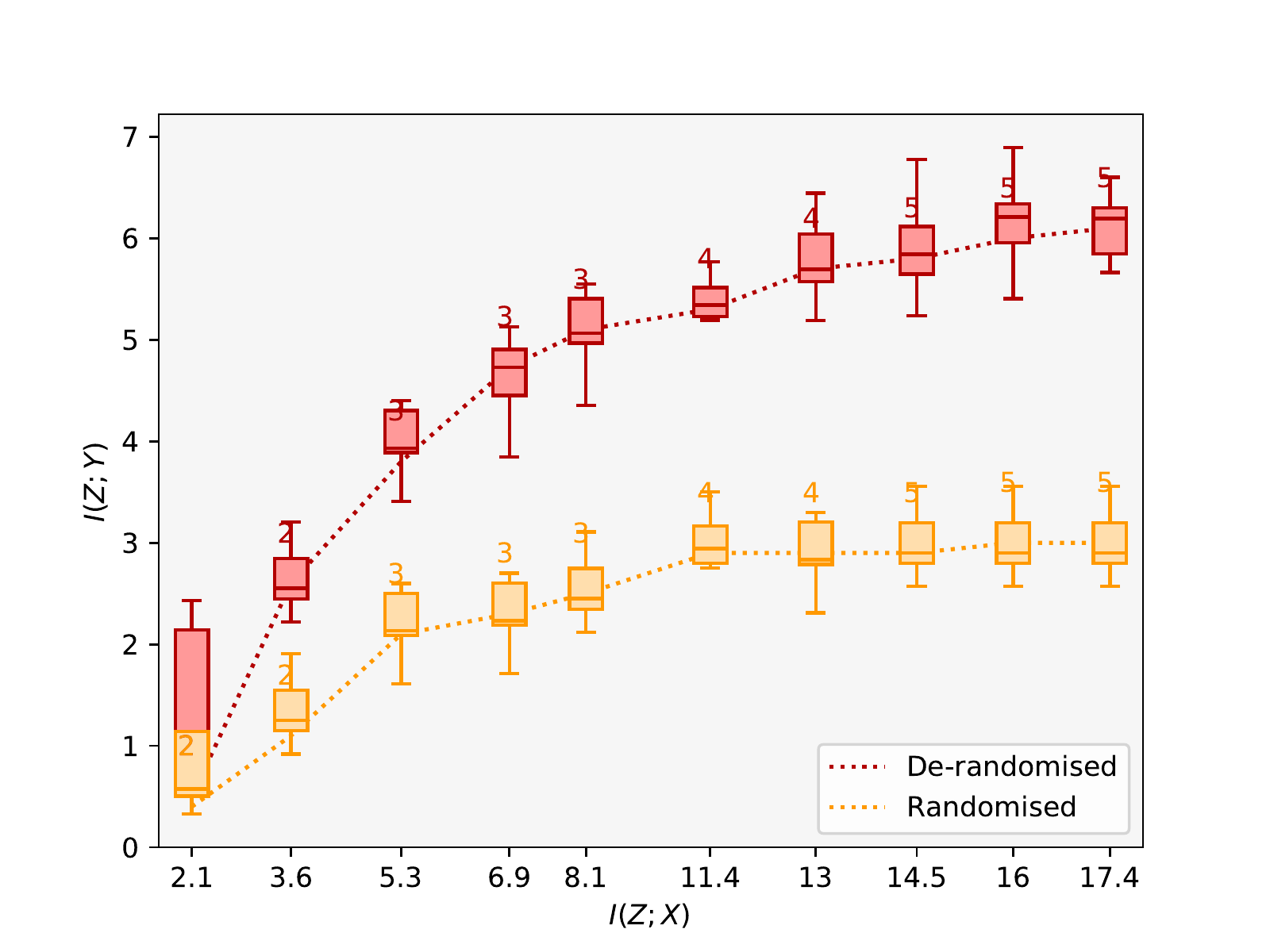}
        		\caption{}
        		\label{information curve:a}
        \end{subfigure}  
    \caption{(a) Information curves for $I(Z; T)$ and (b) $I(Z;Y)$ with de-randomised and randomised data respectively. When the data is randomised, the value of $I(Z;T)$ is  close to zero. The differences between the curves illustrates confounding. When data is de-randomised, we are able to estimate treatment effects more accurately by accounting for this confounding.}  
\end{figure}

To assess the interpretability of the proposed approach and the ability to account for hidden confounding, we perform an analysis on the latent space of our model. First, we plot two information curves illustrating the number of latent dimensions required to reconstruct the output for the terms $I(Z; Y)$ and $I(Z; T)$ respectively. These results are shown in Figure \ref{information curve:b} and Figure \ref{information curve:a}. In particular, we perform this analysis when the data set of subjects is both de-randomised and randomised (i.e. when we do not introduce selection bias into the data set).  Comparing the information curves in Figure \ref{information curve:b} confirms that when we do not de-randomise the data, the information content in the treatment $I(Z; T)$ tends to be closer to 0, whereas the opposite is true when the data is de-randomised. The information curves in Figure \ref{information curve:a} additionally demonstrate our model's ability to account for indirect effects of confounding when predicting the overall outcomes: when data is de-randomised, we are able to reconstruct treatment outcomes more accurately. Overall, the results from Figures \ref{information curve:b} and \ref{information curve:a} highlight that there is indeed a hidden confounding effect that we can account for using the proposed approach.

\begin{figure}[!h]
    \centering
        \begin{subfigure}{0.21\textwidth}
        		\centering
        		\includegraphics[width=\textwidth]{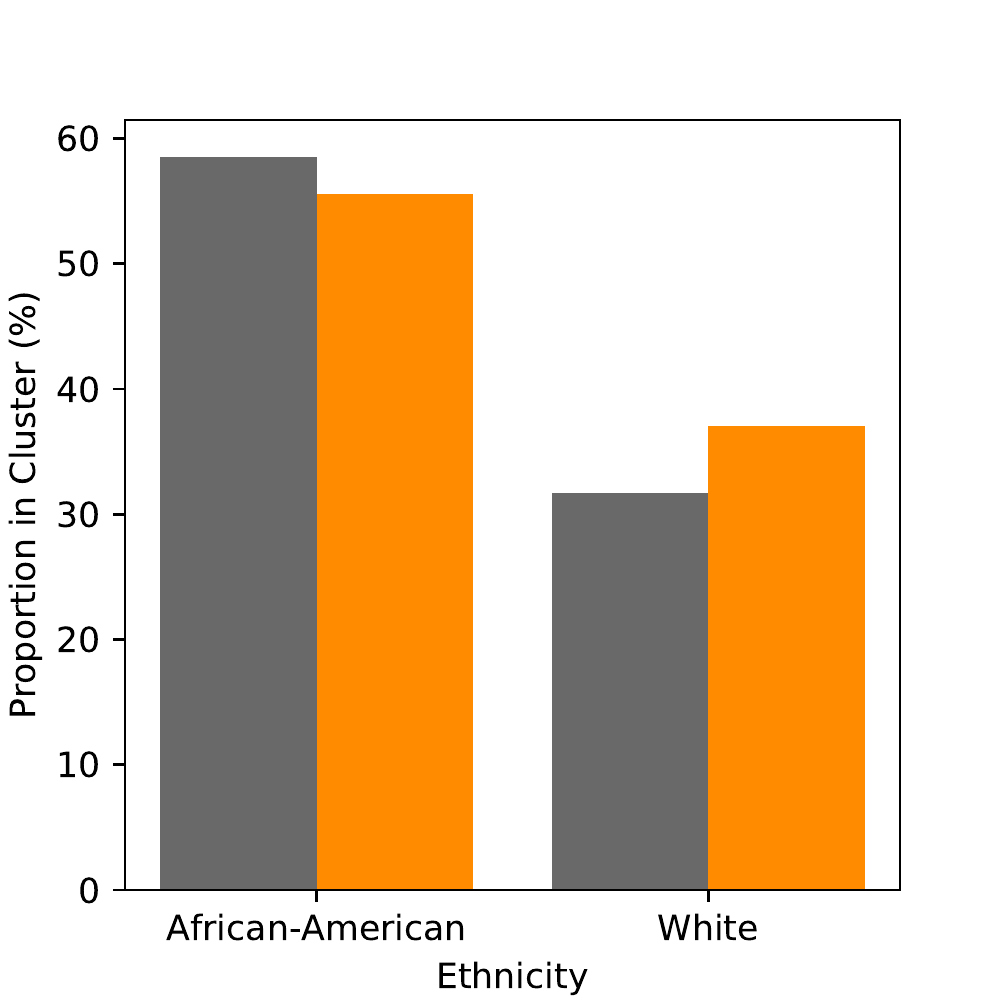}
        		\caption{Cluster 1}
        		\label{cluster1}
        \end{subfigure}         
        \begin{subfigure}{0.21\textwidth}
        		\centering
        		\includegraphics[width=\textwidth]{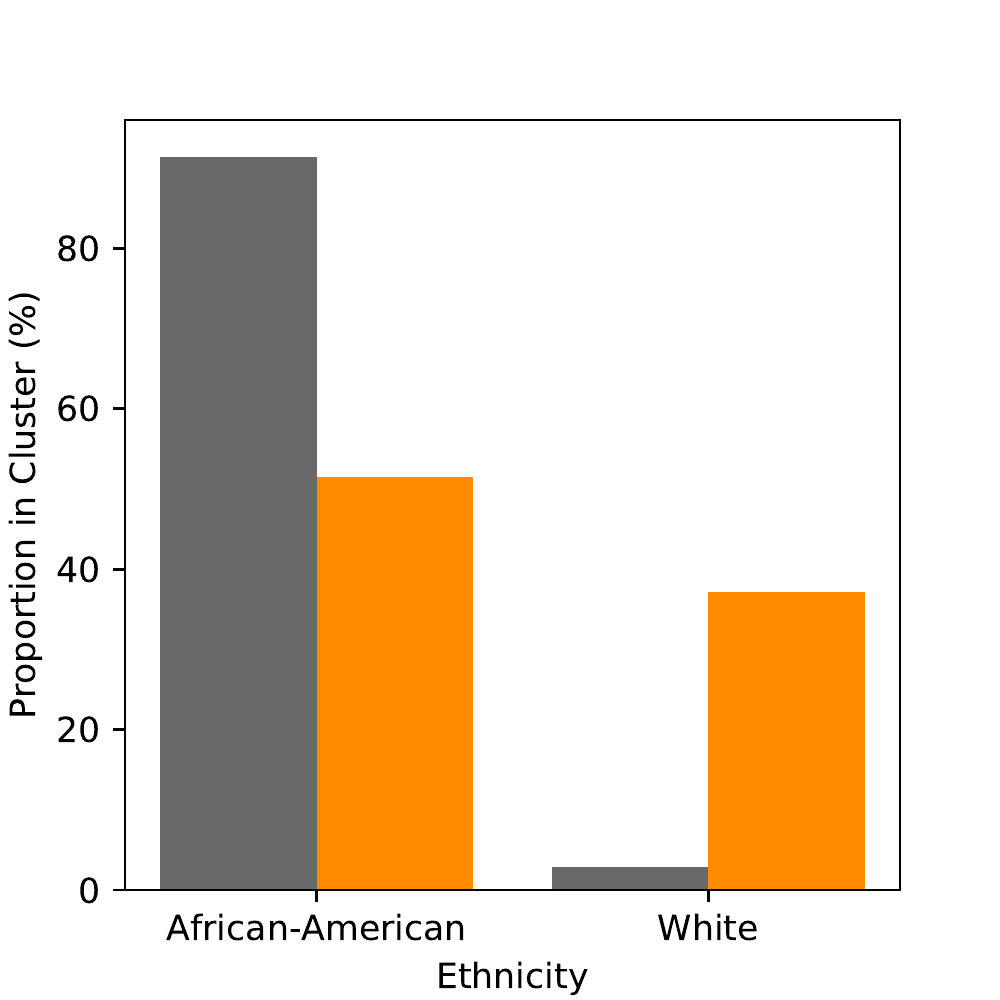}
        		\caption{Cluster 2}
        		\label{cluster2}
        \end{subfigure}
        \begin{subfigure}{0.21\textwidth}
                \centering
        		\includegraphics[width=\textwidth]{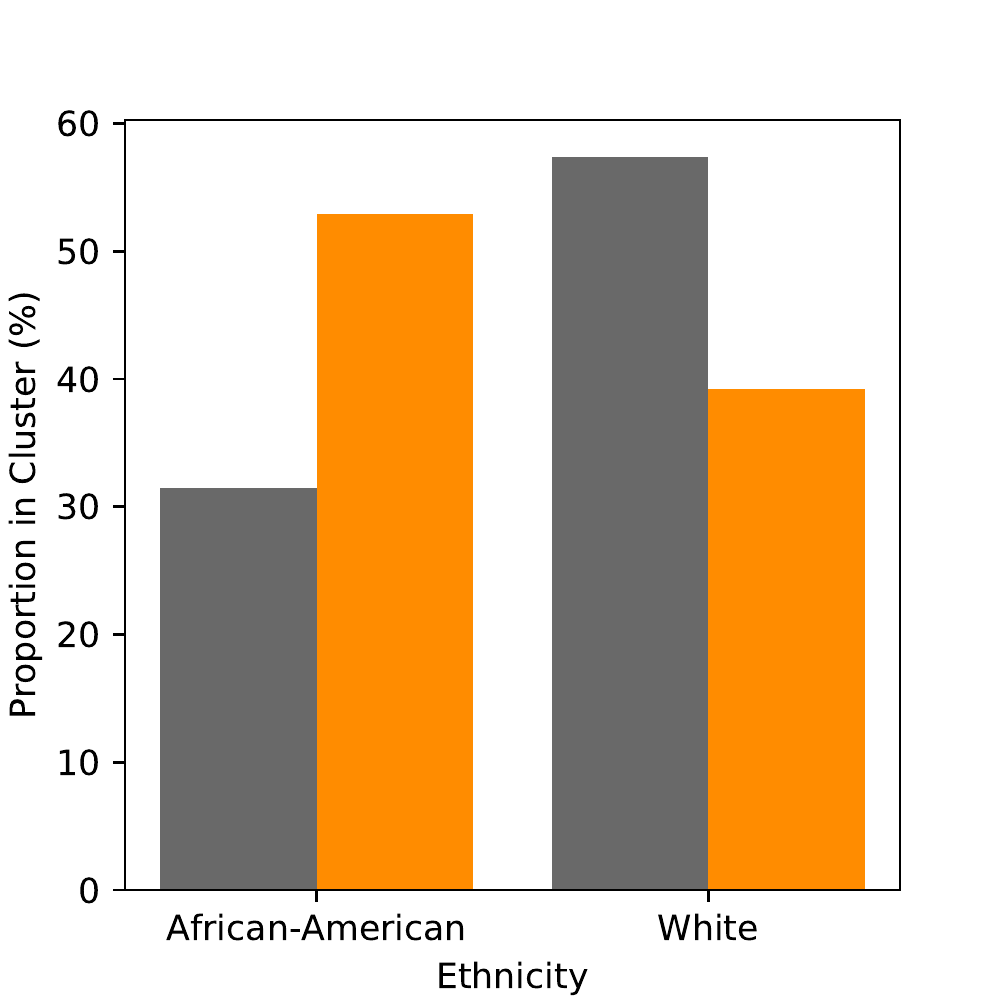}
        		\caption{Cluster 3}
        		\label{cluster3}
        \end{subfigure}  
             \begin{subfigure}{0.21\textwidth}
                \centering
        		\includegraphics[width=\textwidth]{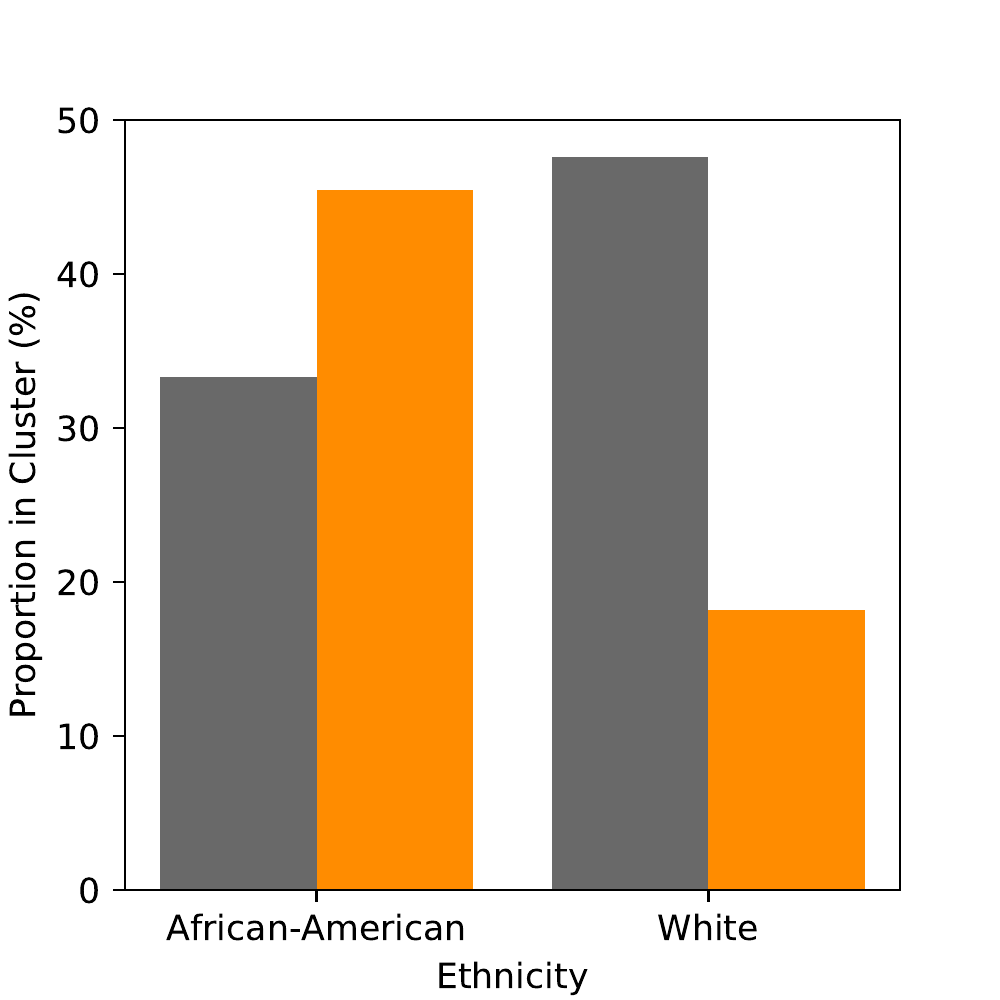}
        		\caption{Cluster 4}
        		\label{cluster4}
        \end{subfigure}  
    \caption{\label{fig:cluster} Illustration of the proportion of major ethnic groups within the four clusters. Grey and orange indicate de-randomised and randomised data respectively. The first cluster in (a) is a neutral cluster. The second cluster in (b) shows an enrichment of information in the African-American group. Clusters 3 and 4 in (c) and (d) respectively, show an enrichment of information in the White group. Overall, we are able to identify the hidden confounder correctly and account for this when predicting outcomes. For better visualisation, we only report the two main clusters which include the majority of all patients.}
\end{figure}

\begin{figure}[!h]
\begin{center}
\includegraphics[width=0.4\textwidth]{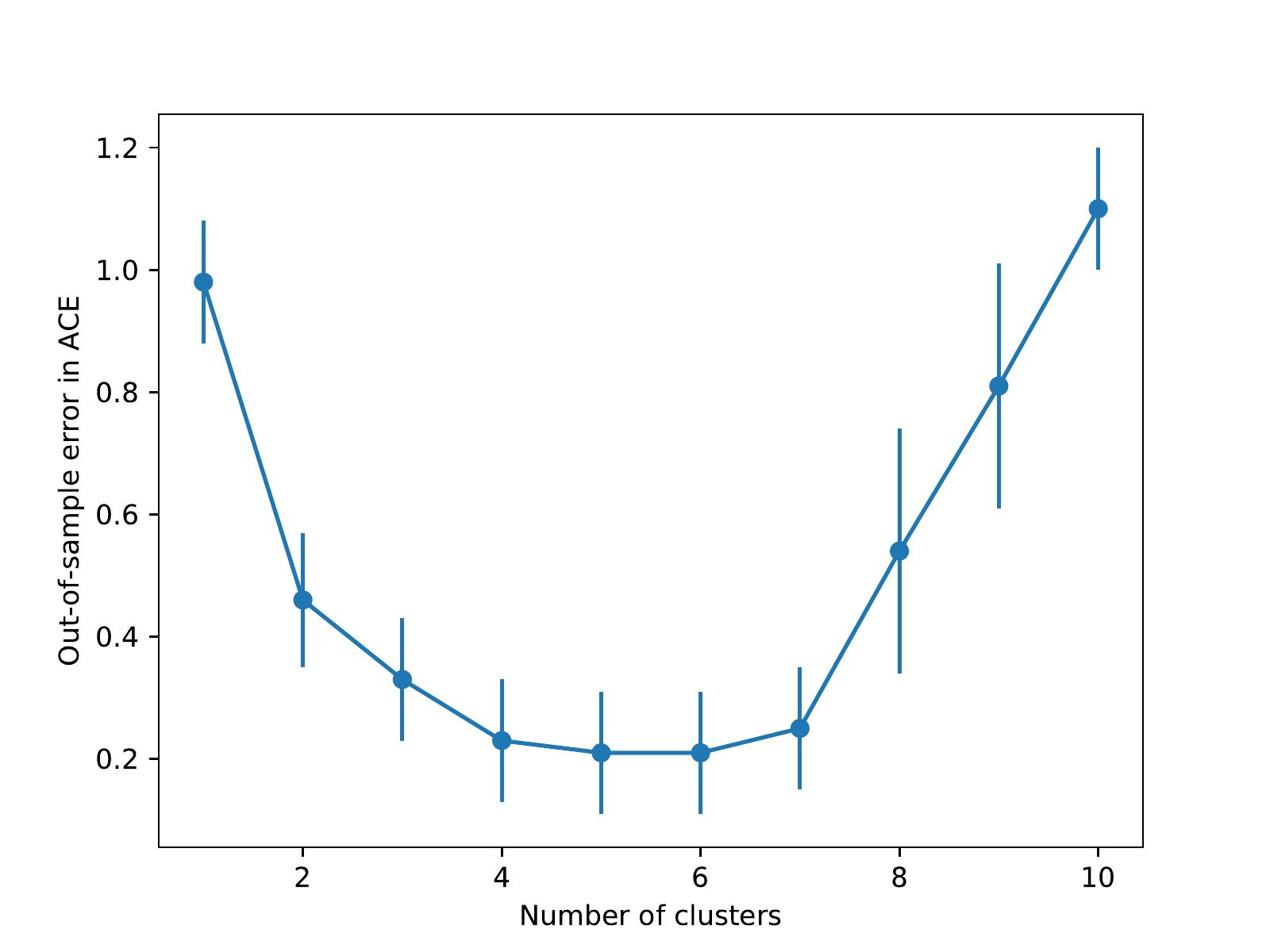}
\caption{Out-of-sample error in ACE with a varying number of clusters.} 
\label{fig:ATE-error-cluster}
\end{center}
\end{figure}

Next, we perform an analysis of the discretised latent space by comparing the proportions of ethnic groups of test subjects in each cluster from the Gaussian mixture to see if we can recover the hidden confounding effect. These results are shown in Figure \ref{fig:cluster} where we plot a hard assignment of test subjects to clusters on the basis of their ethnicity. Evidently, the clusters exhibit a clear structure with respect to the ethnic groups. In particular, Cluster 2 in Figure \ref{cluster2} has a significantly higher proportion of non-white members in the de-randomised setting, confirming that we are able to correctly identify the true confounding effect and account for this when making predictions.  Finally, we perform similar analyses and assess the error in estimating the ACE when varying the number of mixture components in Figure \ref{fig:ATE-error-cluster}.  When the number of clusters is larger, the clusters get smaller and it becomes more difficult to reliably estimate the ACE since we average over the cluster members to account for partial covariate information at test time. Here, model selection is made by observing where the error in estimating the ACE stabilises (anywhere between 4-7 mixture components).


\subsection{Binary Treatment Outcome on Twins}
Like \citet{Louizos}, we apply CEIB to a benchmark task using the birth data of twins in the USA between 1989 and 1991 \citep{almond2005costs}. Here, treatment $T = 1$ is a binary indicator of being the heavier twin at birth, while outcome $Y$ corresponds to the mortality within a year after birth. Since mortality is rare, we consider only same sex twins with weights less than 2 kg which results in 11\,984 pairs of twins. Each twin has a set of 46 covariates including information about their parents such as their level of education, race, incidence of renal disease, diabetes, smoking etc. as well as whether the birth took place in hospital or at home and the number of gestation weeks prior to birth.

\begin{figure}[!h]
\begin{center}
\includegraphics[scale=0.58]{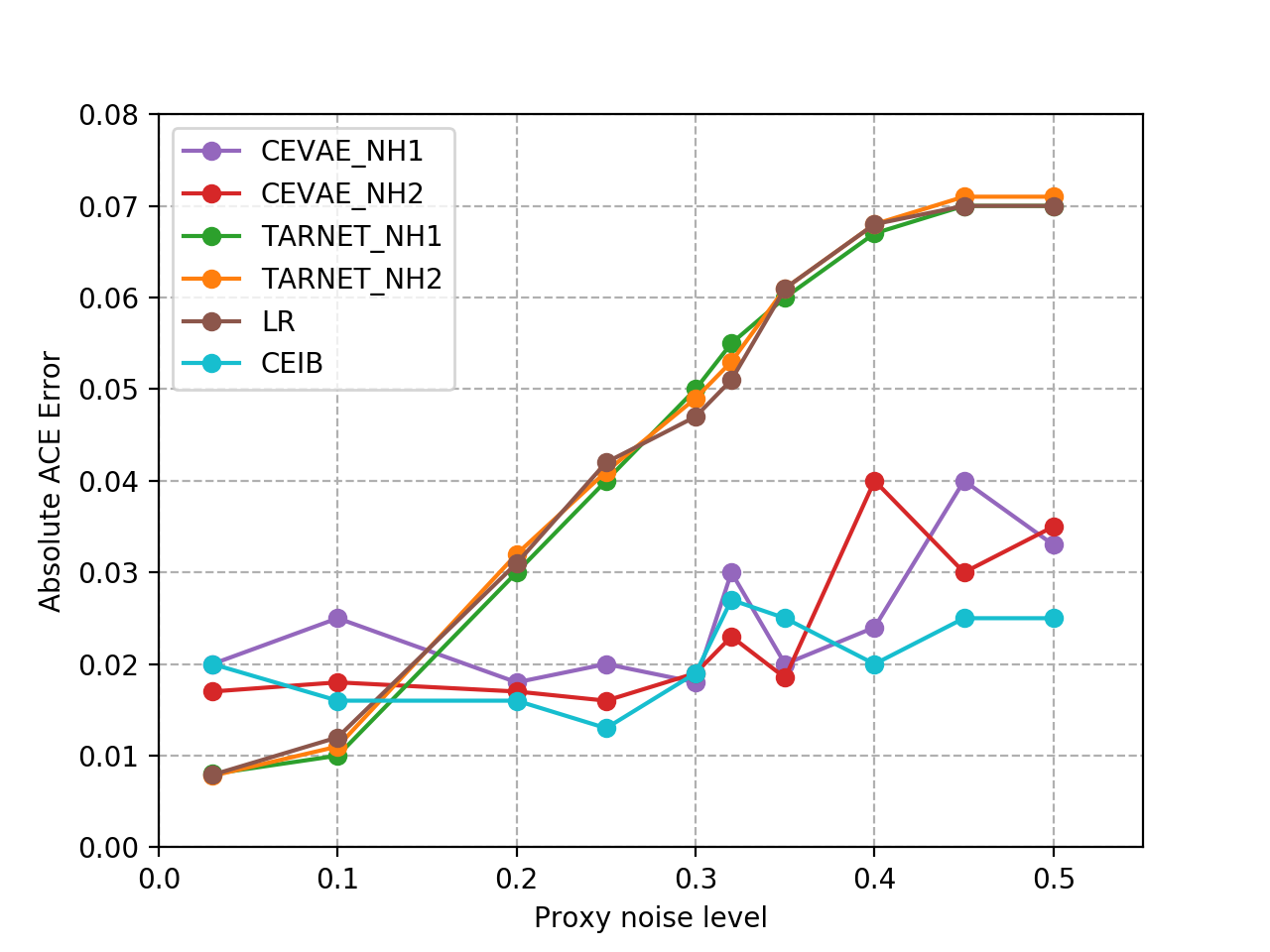}
\caption{Absolute error in ACE estimation for Twins task. CEIB outperforms baselines over varying levels of proxy noise. }
\label{fig: ace-twins}
\end{center}
\end{figure}

To simulate an observational study, we selectively hide one of the twins. To illustrate the ability of CEIB to be applied to Case II where we treat proxy variables as measured confounders, we base the treatment assignment on a single variable which is highly correlated with the outcome: GESTAT10, the number of gestation weeks prior to birth. This has values from 0-9 that correspond to the weeks of gestation before birth i.e. birth before 20 weeks gestation, 20-27 weeks of gestation, etc. Analogous to \cite{Louizos} we set treatment to $t|x, z \sim \texttt{Bern}(\sigma(w_o^\top x + w_h(z/10 -0.1)))$ for $w_o \sim \mathcal{N} (0, 0.1I), w_h \sim \mathcal{N}(5, 0.1)$, where $z$ is GESTAT10 and $x$ are the 45 remaining covariates. Since CEIB can account for incomplete covariates, we artificially exclude 3 covariates from $x$ at test time.

Like \citet{Louizos}, proxies are created with a one-hot encoding of $z$, replicated 3 times and randomly flipping the 30 bits, where the flipping probability varies from 0.05 to 0.15. There may also be additional proxy variables for $z$ in the data from the set of variables. Our task is to predict the ACE. Specifically, we compare the performance of CEIB to CEVAE (with a varying number of hidden layers), TARnet (with varying numbers of hidden layers) and logistic regression (LR). These results are shown in Figure \ref{fig: ace-twins}. Here too, CEIB achieves close to state-of-the-art performance on the Twins task.

\end{document}